\documentclass[preprint,3p,sort&compress,12pt]{elsarticle}

\usepackage{amssymb}
\usepackage{amsmath}
\usepackage{float}
\usepackage{booktabs} 

\begin{document}


\title{A Navier-Stokes-Peridynamics hybrid  algorithm for the coupling of compressible flows and fracturing materials}

\author[CAS,UCAS]{Mingshuo Han\fnref{fn1}}
\author[CAS]{Shiwei Hu\fnref{fn1}}
\author[CAS,UCAS]{Tianbai Xiao\corref{cor1}}
\author[CAS,UCAS]{Yonghao Zhang\corref{cor1}}

\cortext[cor1]{Corresponding authors: txiao@imech.ac.cn, yonghao.zhang@imech.ac.cn}
\fntext[fn1]{These authors contributed equally to this work.}

\affiliation[CAS]{organization={Centre for Interdisciplinary Research in Fluids, Institute of Mechanics, Chinese Academy of Sciences},
            city={Beijing},
            postcode={100190}, 
            country={China}}            

\affiliation[UCAS]{organization={School of Engineering Science, University of Chinese Academy of Sciences},
            city={Beijing},
            postcode={100049}, 
            country={China}} 

\begin{abstract}
Modeling and simulation of fluid-structure interactions are crucial to the success of aerospace engineering.
This work addresses a novel hybrid algorithm that models the close coupling between compressible flows and deformable materials using a mesoscopic approach.
Specifically, the high-speed flows are described by the gas-kinetic scheme, which is a robust Navier-Stokes alternative solver built on the molecular kinetic theory.
The deformation, damage, and fracture of materials are depicted using the bond-based peridynamics, which serves as coarse-grained molecular dynamics to construct non-local extensions of classical continuum mechanics.
The evolution of fluids and materials are closely coupled using the ghost-cell immersed boundary method.
Within each time step, the solutions of flow and solid fields are updated simultaneously, and  physics-driven boundary conditions are exchanged for each other via ghost cells.
Extensive numerical experiments, including crack propagation in a pre-cracked plate, subsonic flow around the NACA0012 airfoil, supersonic flow around the circular cylinder, and shock wave impacting on the elastic panel, are performed to validate the algorithm.
The simulation results demonstrate the unique advantages of current hybrid algorithm in solving fracture propagation induced by high-speed flows.
\end{abstract}




\maketitle

\begin{table}
    \centering
    \caption{Nomenclature.}
    \begin{tabular*}{14cm}{lll}
        \hline
        FSI & fluid-structure interaction \\
        GKS & gas-kinetic scheme \\
        BBPD & bond-based Peridynamics \\
        $t$, $\mathbf x$ & time and space variables \\
        $\mathbf v$, $\zeta$ & particle velocity and internal energy variables \\
        $f$ & particle distribution function \\
        $\mathcal Q$ & collision operator in the kinetic equation \\
        $g$ & Maxwellian equilibrium distribution function \\
        $\tau$ & relaxation time \\
        $\rho$, $\mathbf V$, $T$, $\lambda$ & primitive flow variables \\
        $\mathbf P$, $\mathbf q$ & stress tensor and heat flux \\
        $\psi$ & collision invariants \\
        $\mathbf W$ & conservative flow variables \\
        $\rho_s$ & solid density \\
        $\mathbf u$ & displacement \\
        $\delta$ & horizon radius \\
        $\mathbf{f}$ & force density vector \\
        $\mathbf{b}$ & body density vector \\
        $\mathbf{\xi}, \mathbf{\eta}$ & relative position and displacement vectors \\
        $\omega$ & micropotential \\
        $s$ & bond stretch \\
        $c$ & micromodule function \\
        $\bar{\mathbf W}$ & cell-averaged conservative variables \\
        $\mathbf F^W$ & numerical fluxes for conservative variables \\
        $\Delta \mathbf S$ & area vector of cell faces \\
        $g_0$, $f_0$ & initial equilibrium and non-equilibrium distributions in GKS \\
        $a$, $b$, $A$ & normal, tangetial, and temporal derivatives in GKS \\
        $\nu$, $\beta$ & volume and surface effect correction factors \\
        GC & ghost cell \\
        IP & imaginary point \\
        BI & boundary intersection \\
        $\mathbf{\mathrm{n}}_s$ & outward unit normal vector at boundary \\
        \hline
    \end{tabular*}
    \label{table:nomenclature}
\end{table}

\section{Introduction}
\label{sec1}

The impact of multi-physics coupling on aerodynamic performance becomes increasingly significant on long-endurance and high-maneuverability aircraft \cite{sziroczak2016review}.
Among others, fluid-structure interaction (FSI) is a typical example of multi-physics coupling problems \cite{bak2}.
As the speed and altitude of aircraft increase, the enhanced compressibility leads to intricate flow structures, which causes difficulties in calculation of the stress and heat flux acted on the structure.
The irregular aerodynamic forcing and heating loads can cause deformation, damage, fracture and failure of the material of the vehicle, which in turn inversely affect the flow evolution.
Such coupling effects often emerge in an extremely nonlinear manner under the flight condition.
Accurate modeling and simulation of fluid-structure interactions are the key to the success of aerospace engineering \cite{bak3}.



To simulate FSI problems,
a common strategy is to combine the numerical methods, e.g., the finite volume method (FVM) for computational fluid dynamics (CFD) and the finite element method (FEM) for computational solid mechanics, through effective message passing.
Such coupling can be performed in two methods.
The first idea is to generate two sets of body-fitted meshes that are capable of Lagrangian movement and deformation, so that the location change of fluid-structure interface can be tracked \cite{souli2013arbitrary,tezduyar2008interface}.
The second approach is to model the boundary effects on a uniform Eulerian mesh, i.e., the immersed boundary method (IBM) \cite{verzicco2023immersed,sotiropoulos2014immersed,RN13}.
It has been widely regarded for the ease of implementation and flexibility in handling complex boundaries.
Related studies have been carried out on rigid-body \cite{kallemov2016immersed,liao2010simulating,kim2016penalty,su2007immersed} and flexible-body dynamics \cite{huang2011improved,gilmanov2005hybrid,favier2014lattice,lee2015discrete} coupled with  incompressible \cite{choi2007immersed,RN4,RN7,azis2019immersed} and compressible flows \cite{RN14,RN15,RN10,ghias2007sharp,com2,com3}.

Despite the success of IBM-based research on FSI problems, studies on structural failure under external flows are still scarce.
A main cause for this phenomenon is the limitation of traditional FEM in handling fracture problems.
Based on the partial differential equations in the continuum mechanics, the FEM model suffers from ill-defined derivatives around spatial discontinuities, e.g., fractures or defects in the computational domain.
Remedial techniques have been proposed to address this challenge, including remeshing technique that conform the mesh to discontinuities dynamically \cite{shahani2009finite,ramalho2021novel} and the extended finite element method (XFEM) that capture the behavior of discontinuities within elements by enriching the solution space with additional functions \cite{RN16, RN17}.
Although these methods extend the capability of FEM to describe discontinuities solution, they also lead to a number of problems in terms of computational overhead, degraded accuracy, and handling multiple discontinuities.

Peridynamics (PD) is a systematic theory proposed for fracture mechanics \cite{RN18}.
Based on the material point description, it models, point by point, the accumulative effects of its interactions with all surrounding points within a certain range.
Therefore, the PD theory replaces the local interactions described by differential equations in classical continuum mechanics with non-local interactions described by integral equations, thus circumventing the challenges posed by spatial discontinuities and non-differentiable fields. 
When the relative displacement between two material points exceeds a critical threshold, their mutual interactions disappear, allowing spontaneous fracture modeling.
This computational framework provides a solid basis for modeling fracture generation and propagation as well as multi-crack interactions and branching phenomena \cite{RN20,zy,RN23,RN24,RN25}.
With two decades of development, the PD theory and related computational methods have become an important alternative to FEM in fracture mechanics \cite{RN26,RN27,RN28,RN29,RN30,RN31,RN32,hu2024efficient}.

Efforts have been initiated to simulate FSI problems using peridynamics.
Gao et al. combined peridynamics for structural deformation and peridynamic differential operators (PDDO) for fluid motion to investigate the interaction between weakly compressible fluids and structures, and a collision model is built to model the fluid-structure interface as a reflective boundary \cite{RN33}.
Behzadinasab et al. developed a coupled method integrating iso-geometric analysis (IGA) with PD to study the interaction between blast-induced airflow and structures \cite{RN34,RN35}. 
Zhang et al. proposed a strongly coupled method that combines the lattice Boltzmann method (LBM) and PD to solve the incompressible flow-induced structural deformation and fracture \cite{RN36}.
Federico et al. established a coupling approach integrating IBM with ordinary state-based PD to investigate three-dimensional incompressible flow-induced structural deformation and fracture \cite{RN37}.
Kim et al. introduced an immersed peridynamics method based on the direct-forcing IBM to study large deformation and failure of hyperelastic materials under incompressible fluid flow \cite{RN38}.

Current FSI studies based on peridynamics are mainly focused on incompressible and weakly compressible flows.
As the flow velocity increases, the effects of compressibility become increasingly significant.
The evolution of the wave system, including shock and rarefaction waves, makes the mechanism of FSI more complex.
This work addresses this gap by developing a closely coupled computational framework.
It combines the gas-kinetic scheme for compressible Navier-Stokes flows \cite{xukun,chen2016cartesian,xiao2020velocity} and the bond-based peridynamics for fracturable materials.
The evolution of fluid and structure is coupled using the ghost-cell immersed boundary method.
The developed framework is capable of high-fidelity and real-time simulation of flow-induced structural deformation and fracture.




The rest of the paper is structured as follows. 
Section \ref{sec2} briefly introduces the basic theory of fluids and solids.
Section \ref{sec3} presents the formulation of the solution algorithm and its detailed implementations.
Section \ref{sec4} includes numerical experiments to demonstrate the performance of the current method.
The last section presents concluding remarks.
The nomenclature is presented in Table \ref{table:nomenclature}.

\section{Theory of Fluids and Solids}\label{sec2}
\subsection{Gas kinetic theory}


Gas kinetic theory provides an insightful perspective for understanding non-equilibrium fluid transport and designing numerical schemes that are compatible with entropy-increasing processes.
The Boltzmann equation and its simplified models describe the evolution of a many-particle system statistically based on the single-particle distribution function.
The Boltzmann-BGK model \cite{bhatnagar1954model} takes the form
\begin{equation}
\partial_t f + \mathbf v \cdot \nabla_{\mathbf x} f = \mathcal Q(f) = \frac{g - f}{\tau},
\label{eq:bgk}
\end{equation}
where $f(t,\mathbf x,\mathbf v,\zeta)$ is the particle distribution function with respect to time, space, velocity, and internal energy modes. 
The collision operator is denoted by $\mathcal Q$, where $\tau$ is the collision time and $g$ is the Maxwellian that corresponds to thermodynamic equilibrium state, i.e.
\begin{equation}
g = \rho \left( \frac{\lambda}{\pi} \right)^{\frac{N+3}{2}} \exp \left[ -\lambda \left( (\mathbf v - \mathbf V)^2 + \zeta^2 \right) \right],
\label{eq:maxwellian}
\end{equation}
where $\rho$ is the density, $\mathbf V$ is the macroscopic velocity, and $\lambda = m / 2kT$, with $m$ being the molecular mass, $k$ the Boltzmann constant, and $T$ the temperature. 
The degrees of freedom of internal energy takes $(5 - 3\gamma)/(\gamma - 1) + 1$, where $\gamma$ is the adiabatic index.
The internal variable in the equilibrium is thus given by $\zeta^2 = \zeta_1^2 + \dots + \zeta_N^2$.
Note that the BGK model provides the fixed Prandtl number $\mathrm{Pr}=1$.
This deficiency can be ameliorated by correcting the equilibrium distribution in Eq.(\ref{eq:bgk}), e.g., the Shakhov \cite{shakhov1968generalization} and ES-BGK models \cite{holway1966new}.

The conservative variables can be derived by taking the moments of the distribution function over the  velocity space, i.e.,
\begin{equation}
\mathbf W=
\begin{pmatrix}
\rho \\
\rho \mathbf V \\
\rho E
\end{pmatrix}
=
\int \psi f \, d {\Xi},
\label{eq:moments}
\end{equation}
where $E$ denotes the total energy density, $\psi$ is the vector of collision invariants,
\begin{equation}
\psi = 
\begin{pmatrix}
1, \mathbf v, \frac{1}{2}(\mathbf v^2 + \zeta^2)
\end{pmatrix}^T,
\label{eq:psi}
\end{equation}
and $d\Xi = d\mathbf v d \zeta$.

It is known that the Euler and Navier-Stokes equations can be derived as the limits of solutions of the Boltzmann equation.
Considering the first-order Chapman-Enskog expansion of $f$, i.e.,
\begin{equation}
    f = g+f_{1} = g - \tau (\partial_t g + \mathbf v \cdot \nabla_{\mathbf x}g),
    \label{eq:ce_expansion}
\end{equation}
we can take conservative moments of the BGK equation, which yields
\begin{equation}
\int \psi (\partial_t g + \mathbf v \cdot \nabla_{\mathbf x} g) d {\Xi} = \tau \int \psi (\partial_t f_{1} + \mathbf v \cdot \nabla_{\mathbf x} f_1) d {\Xi},
\end{equation}
from which the Navier-Stokes equations can be obtained, i.e.,
\begin{equation}
\begin{aligned}
    &\frac{\partial \rho}{\partial t} + \nabla_{\mathbf x} \cdot (\rho \mathbf V)=0, \\
    &\frac{\partial \rho \mathbf V}{\partial t} + \nabla_{\mathbf x} \cdot (\rho \mathbf V \otimes \mathbf V)-\nabla_{\mathbf x} \cdot \mathbf P=0, \\
    &\frac{\partial \rho E}{\partial t} + \nabla_{\mathbf x} \cdot (\rho E\mathbf V)-\nabla_{\mathbf x} \cdot (\mathbf P \cdot \mathbf V)+\nabla_{\mathbf x} \cdot \mathbf q=0,
\end{aligned}
\label{eq:conservation laws}
\end{equation}
where $\otimes$ denotes dyadic product.
The stress tensor and heat flux are defined as
\begin{equation}
    \mathbf P=\int_{\mathbb R^3} (\mathbf v-\mathbf V)\otimes(\mathbf v-\mathbf V)fd\mathbf v, \quad \mathbf q=\int_{\mathbb R^3} \frac{1}{2}(\mathbf v-\mathbf V)(\mathbf v-\mathbf V)^2fd\mathbf v.
\end{equation}
Inserting Eq.(\ref{eq:ce_expansion}) into the above equation gives the Newton's stress formula and Fourier's law, which naturally recovers the Navier-Stokes equations.

In the context of viscosity-dominated flows, the collision time $\tau$ is linked to the dynamic viscosity $\mu$ through the relation $\tau = \frac{\mu}{p}$. Conversely, for heat-conduction-dominated flows, the collision time is selected as $\tau=\frac{\kappa}{c_{p}p}$ to recover the heat conduction coefficient $\kappa$.

The Chapman-Enskog expansion provides a comprehensive and consistent description of equilibrium (inviscid Euler) and non-equilibrium (viscous) effects in the Navier-Stokes equations.
No artificial assumptions other than the truncation order are made.
A family of gas-kinetic schemes (GKS) \cite{xu2014direct,xiao2021flux,xiao2021kinetic} have been developed based on the evolutionary Chapman-Enskog solution within the framework of BGK equation.
In this work, the GKS solver is employed to solve the evolution of compressible flows, see section \ref{sec3.1} for details.



\subsection{Peridynamic theory}

\begin{figure}
\centering
\includegraphics[width=0.6\textwidth]{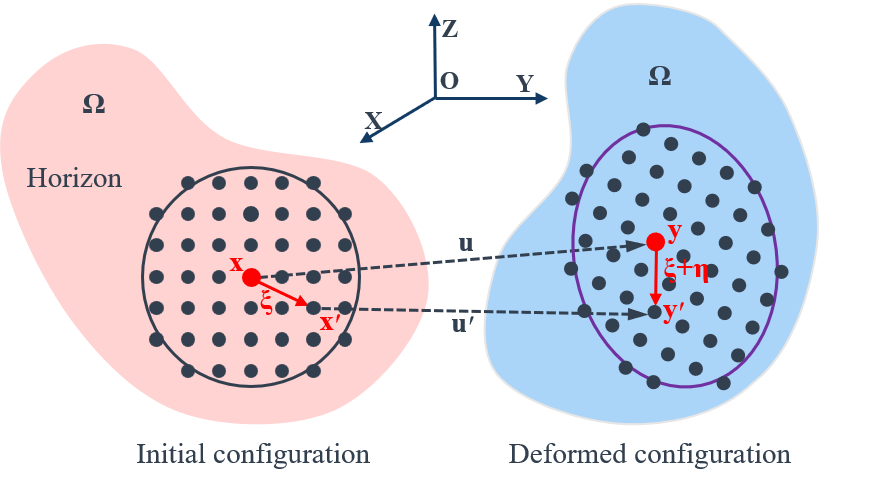}
\caption{Illustration of BBPD and pairwise interaction between PD points. The horizon of the particle at the position $\mathbf{x}$ encompasses all the PD particles interacting with it, and remains unchanged throughout deformation.}\label{fig1}
\end{figure}

Peridynamics is explicitly formulated based on Newton's second law of motion.
The governing equation of bond-based Peridynamic (BBPD) can be written as
%
\begin{equation}\label{pd}
    \rho_s(\mathbf{x})\ddot{\mathbf{u}}(t,\mathbf{x}) = \int_{H_\mathbf{x}}\mathbf{f}(\mathbf{u}^\prime - \mathbf{u},
    \mathbf{x}^\prime - \mathbf{x})d{V^\prime} + \mathbf{b}(t,\mathbf{x}),
\end{equation}
where $\rho_s(\mathbf{x})$ denotes the mass density of solid at the position $\mathbf{x}$.
The displacement is denoted as $\mathbf u$, and thus $\ddot{\mathbf{u}}$ is the acceleration vector field.
The force density vector between the particles $\mathbf{f}$ is a function of the relative position $\mathbf{x'} - \mathbf{x}$ and the relative displacement $\mathbf{u'} - \mathbf{u}$.
The body force density vector is denoted as $\mathbf{b}$.
The interactions between particles are modeled through integrals over the domain of influence. 
The integral domain (also known as horizon) is defined,
\begin{equation}
    H_\mathbf{x} = \{\mathbf{x}' \in \mathbb{R} : |\mathbf{x} - \mathbf{x}'| < \delta\},
\end{equation}
where $\delta$ is the radius of horizon.
The volume occupied by the particle at $\mathbf{x'}$ is denoted as $V^\prime$. 
After the configuration deformation, we denote the new positions of the particles as $\mathbf{y'}$ and $\mathbf{y}$, where $\mathbf{y'}=\mathbf{x'}+\mathbf{u'}$ and  $\mathbf{y}=\mathbf{x}+\mathbf{u}$.  As shown in Figure \ref{fig1}, we introduce $\mathbf{\xi}$ and $\mathbf{\eta}$ to represent the relative position and displacement  vectors between the particles  $\mathbf{x}$ and $\mathbf{x'}$, where $\mathbf{\xi} = \mathbf{x'}-\mathbf{x}$ and $\mathbf{\eta} = \mathbf{u'}-\mathbf{u}$.

For a microelastic material\cite{RN18}, a pairwise potential $\omega$ can be modeled such that
\begin{equation}\label{new1}
    \mathbf{f}(\mathbf{\eta},\mathbf{\xi}) = \frac{\partial\omega(\mathbf{\eta}, \mathbf{\xi})}{\partial\mathbf{\eta}},\quad \forall \mathbf{\xi}, \mathbf{\eta}.
\end{equation}
Under simple linear microelasticity, the micropotential is defined as
\begin{equation}\label{new2}
    \omega(\mathbf{\eta}, \mathbf{\xi}) =
    \frac{cs^2\|\mathbf{\xi}\|}{2},
\end{equation}
where $c$ represents the micromodule function, and $s$ represents the bond stretch, i.e.,
\begin{equation}
s = \frac{\| \mathbf{\xi}+\mathbf{\eta}\|-\|\mathbf{\xi}\|}{\|\mathbf{\xi}\|}.
\end{equation}
Therefore, the corresponding pairwise force $\mathbf{f}$ becomes 
\begin{equation}
    \mathbf{f}(\mathbf{\eta},\mathbf{\xi}) =  \frac{\partial\omega(\mathbf{\eta}, \mathbf{\xi})}{\partial\mathbf{\eta}} = cs\|\mathbf{\xi}\|
    \frac{\partial s}{\partial\mathbf{\eta}}=
    cs\frac{\mathbf{\xi}+\mathbf{\eta}}{\|\mathbf{\xi}+\mathbf{\eta}\|}.
\end{equation}
Here, the micromodule function $c$ takes usually the constant form as
\begin{equation}
c = \begin{cases} 
\frac{2E}{\pi\delta^2A}, & \text{for 1D,} \\
\frac{9E}{\pi\delta^3h}, & \text{for plane stress,} \\
\frac{48E}{5\pi\delta^3h}, & \text{for plane strain,} \\
\frac{12E}{\pi\delta^4}, & \text{for 3D.}
\end{cases}
\end{equation}

Failure of functional materials can be naturally formulated in the BBPD since it allows a bond to break when its stretch exceeds a predefined threshold.
The tensile force in the bond is no longer sustained after the failure.
Therefore, we modify the bond force density as
\begin{equation}\label{new3}
    \mathbf{f} = \lambda_s cs\frac{\mathbf{\xi}+\mathbf{\eta}}{\|\mathbf{\xi}+\mathbf{\eta}\|}.
\end{equation}
Here, $\lambda_s$ is a history-dependent scalar valued function that describes the state of bond failure.  When the bond stretch $s$ exceeds the critical bond stretch $s_c$, the bond state are set to zero, i.e.,
\begin{equation}
\lambda_s = \begin{cases} 
1.0, & \text{if } s < s_c, \\ 
0.0, & \text{if } s \geq s_c.
\end{cases}
\end{equation}
The critical bond stretch $s_c$ can be determined by the material's critical energy release rate $G_c$ \cite{crack}, i.e.,
\begin{equation}
s_c =  \begin{cases} 
\sqrt{\frac{\pi G_c}{3k \delta}} & \text{for 2D},\\ 
\sqrt{\frac{5 G_c}{9k \delta}} & \text{for 3D},
\end{cases}
\end{equation}
where $k$ is the bulk modulus of the material.

\section{Solution Algorithm}\label{sec3}
\subsection{Gas-kinetic scheme}\label{sec3.1}

The gas-kinetic scheme is a finite volume method.
The computational domain $\mathbf \Omega$ is divided into $N_{x}$ non-overlapping elements, and the cell-averaged conservative variables are introduced as
\begin{equation}
    \bar{\mathbf W}_{i}=\frac{1}{V_i} \int_{\mathbf \Omega_{i}} \mathbf W(t,\mathbf x) d\mathbf x.
\end{equation}
Integrating Eq.(\ref{eq:conservation laws}) yields the
update algorithm of $\bar{\mathbf W}$, i.e.,
\begin{equation}
\begin{aligned}
    \frac{\partial \bar{\mathbf W}_{i}}{\partial t} &= -\frac{1}{V_{i}} \oint_{\partial \mathbf \Omega_i} \mathbf F^W \cdot d\mathbf S = -\frac{1}{V_{i}} \sum_{k=1}^{N_f} \mathbf F_{k}^W \cdot \Delta \mathbf S_k,
\end{aligned}
\label{eq:fvm macro}
\end{equation}
where $\mathbf F^W$ denotes the numerical fluxes, $\mathbf S=\mathbf n \Delta S$ is the area vector pointing out of the cell, and $N_f$ is the number of faces.

The core task in the finite volume method is to accurately compute the numerical fluxes.
For brevity, we employ the two-dimensional Cartesian coordinate system to introduce the principle of gas-kinetic flux solver.
It can be straightforwardly extended to non-orthogonal mesh using local coordinate transformation.

Under the assumption that the collision time is a local constant, the integral solution of the BGK equation in Eq.(\ref{eq:bgk}) at a cell face $(x_{i+1/2},y_j)$ writes
\begin{equation}
    \begin{aligned}f(x_{i+1/2},y_{j},t,u,v,\zeta)&=\frac{1}{\tau}\int_{0}^{t}g(x^{\prime},y^{\prime},t^{\prime},u,v,\zeta)e^{-(t-t^{\prime})/\tau}dt^{\prime}\\&+e^{-t/\tau}f_{0}(x_{i+1/2}-ut,y_{j}-vt),\end{aligned}
\end{equation}
where $x'=x_{i+1/2}-u(t-t')$, $y'=y_{j}-v(t-t')$ are the particle trajectories, $f_0$ is the initial distribution function at the beginning of each time step $(t=0)$.
Two unknowns $g$ and $f_0$ are constructed to concretize the above equation and construct the numerical fluxes.
We assume, without loss of generality, that \(x_{i + 1/2}=0\) and \(y_j = 0\) to simplify the notation.

The initial distribution function is constructed following the piecewise Chapman-Enskog expansion, i.e.,
\begin{equation}
    f_0=\left\{\begin{aligned}&g^l\left(1+a^lx+b^ly-\tau(a^lu+b^lv+A^l)\right), \ x\leq0 \\
    &g^r\left(1+a^rx+b^ry-\tau(a^ru+b^rv+A^r)\right), \ x>0\end{aligned}\right.
    \label{eq:flux f}
\end{equation}
where $g^l$ and $g^r$ are the Maxwellian distributions related to the left and right-hand states around a face, while $\{a^l,a^r\}$ and $\{b^l,b^r\}$ are related to the slopes in the normal and tangential directions.

The equilibrium state is constructed as
\begin{equation}
    g=g_0(1+(1-H(x))\bar a^l x+H(x)\bar a^r x + \bar b y + \bar A t),
    \label{eq:flux g}
\end{equation}
where $g_0$ is the initial Maxwellian distribution at the face, while $\{\bar a^l,\bar a^r\}$ and $\bar b$ are connected with the normal and tangential slopes of $g$.
The $H(x)$ is the heaviside step function,
\begin{equation}
H(x) = \begin{cases} 
0, & \text{if } x < 0, \\ 
1, & \text{if } x >= 0.
\end{cases}
\end{equation}

In both $f$ and $g$, the slopes are related to the  gradients of macroscopic variables via
\begin{equation}
    \int f a\psi d\Xi=\partial_x \mathbf W, \ \int f b\psi d\Xi=\partial_y \mathbf W,
\end{equation}
and can then be determined by solving a linear system.
Substituting Eqs.(\ref{eq:flux f}) and (\ref{eq:flux g}) yields the time dependent solution at the face,
\begin{equation}
\begin{aligned}
f(x_{i+\frac{1}{2}}, y_j, t, u, v, \xi) = & \ (1 - e^{-t/\tau}) g_0 \\
& + \left[ \tau(-1 + e^{-t/\tau}) + t e^{-t/\tau} \right] 
\left[ \bar{a}^l H(u) + \bar{a}^r (1 - H(u)) \right] u g_0 \\
& + \tau \left( \frac{t}{\tau} - 1 + e^{-t/\tau} \right) \bar{A} g_0 \\
& + e^{-t/\tau} \left[ (1 - u(t + \tau) \bar{a}^l) H(u) g^l 
+ (1 - u(t + \tau) \bar{a}^r) (1 - H(u)) g^r \right] \\
& + e^{-t/\tau} \left[ -\tau A^l H(u) g^l 
- \tau A^r (1 - H(u)) g^r \right].
\end{aligned}
\label{eq:distribution_function}
\end{equation}
The numerical fluxes can thus be computed by
\begin{equation}
\begin{pmatrix}
F_\rho \\
F_{\rho U} \\
F_{\rho V} \\
F_{\rho E}
\end{pmatrix}_{i+\frac{1}{2},j}
=
\int u 
\begin{pmatrix}
1 \\
u \\
v \\
\frac{1}{2}(u^2 + v^2 + \zeta^2)
\end{pmatrix}
f \left( x_{i+\frac{1}{2}}, j, t, u, v, \zeta \right) d\Xi.
\label{eq:flux_gks}
\end{equation}

\subsection{Peridynamic solver}

In the peridynamic solver, the solid region is discretized into a group of material points.
The integral operator in Eq.(\ref{pd}) is approximated by the numerical quadrature.
The semi-discrete governing equation of the bond-based peridynamics (BBPD) writes
\begin{equation}\label{new4}
    \rho_s(\mathbf{x}_i) \ddot{\mathbf{u}}(t,\mathbf{x}_i) = \sum_{j=1}^{N_i} [\mathbf{f}_{i,j}(\mathbf{u}_j - \mathbf{u}_i,
    \mathbf{x}_j - \mathbf{x}_i)
    \beta _{i,j}\nu _{i,j} V_j] + \mathbf{b}(t,\mathbf{x}_i),
\end{equation}
where $N_i$ is the number of particles within the horizon of the particle located at $\mathbf{x}_i$. The subscript $j$ denotes an arbitrary particle positioned at $\mathbf{x}_j$ within the horizon of the particle at $\mathbf{x}_i$, and $V_j$ is the volume of the particle at $\mathbf{x}_j$ within the horizon of the particle at $\mathbf{x}_i$.
For the simplest uniform distribution of points, $V_j = (\Delta x)^3$.
Here, $\nu_{i,j}$ and $\beta_{i,j}$ are the correction factors for the volume and surface effects respectively for the particle at $\mathbf{x}_j$ within the horizon of the particle at $\mathbf{x}_i$ and $\Delta x$ represents the particle spacing.\par
Based on Eq.(\ref{new3}), the  spatially discrete form of the force density vector of each bond $\mathbf{f}_{i,j}$ satisfies 
\begin{equation}\label{new5}
\begin{aligned}
    \mathbf{f}_{i,j}(\mathbf{u}_j - \mathbf{u}_i, \mathbf{x}_j - \mathbf{x}_i) &=  cs\lambda_s \frac{\mathbf{\xi}_{ij} + \mathbf{\eta}_{ij}}{\left| \mathbf{\xi}_{ij} + \mathbf{\eta}_{ij} \right|} \\
    &= c\lambda_s \frac{\left| \mathbf{\xi}_{ij} + \mathbf{\eta}_{ij} \right| - \left| \mathbf{\xi}_{ij} \right|}{\left| \mathbf{\xi}_{ij} \right|} \frac{\mathbf{\xi}_{ij} + \mathbf{\eta}_{ij}}{\left| \mathbf{\xi}_{ij} + \mathbf{\eta}_{ij} \right|} \\
    &= c\lambda_s \frac{\left| \mathbf{y}_j - \mathbf{y}_i \right| - \left| \mathbf{x}_j - \mathbf{x}_i \right|}{\left| \mathbf{x}_j - \mathbf{x}_i \right|} \frac{\mathbf{y}_j - \mathbf{y}_i}{\left| \mathbf{y}_j - \mathbf{y}_i \right|}.
\end{aligned}
\end{equation}

The explicit forward and backward difference scheme is employed as the integrator along the time direction.
The dynamic evolution of displacement, velocity, and acceleration is computed as
\begin{equation}
\begin{aligned}
    &\ddot{\mathbf{u}}_i^{(n)} = \frac{\mathbf{f}_{i,j}^{(n)} + \mathbf{b}^{(n)}(\mathbf{x}_i,n \Delta t)}{\rho(\mathbf{x}_i)}, \\
    &\dot{\mathbf{u}}_i^{(n+1)} = \dot{\mathbf{u}}_i^{(n)} + \ddot{\mathbf{u}}_i^{(n)} \Delta t, \\
    &\mathbf{u}_i^{(n+1)} = \mathbf{u}_i^{(n)} + \dot{\mathbf{u}}_i^{(n+1)} \Delta t.
\end{aligned}
\end{equation}


\subsection{Immersed boundary method}

\begin{figure}[htb!]
    \centering
    \includegraphics[width=0.6\textwidth]{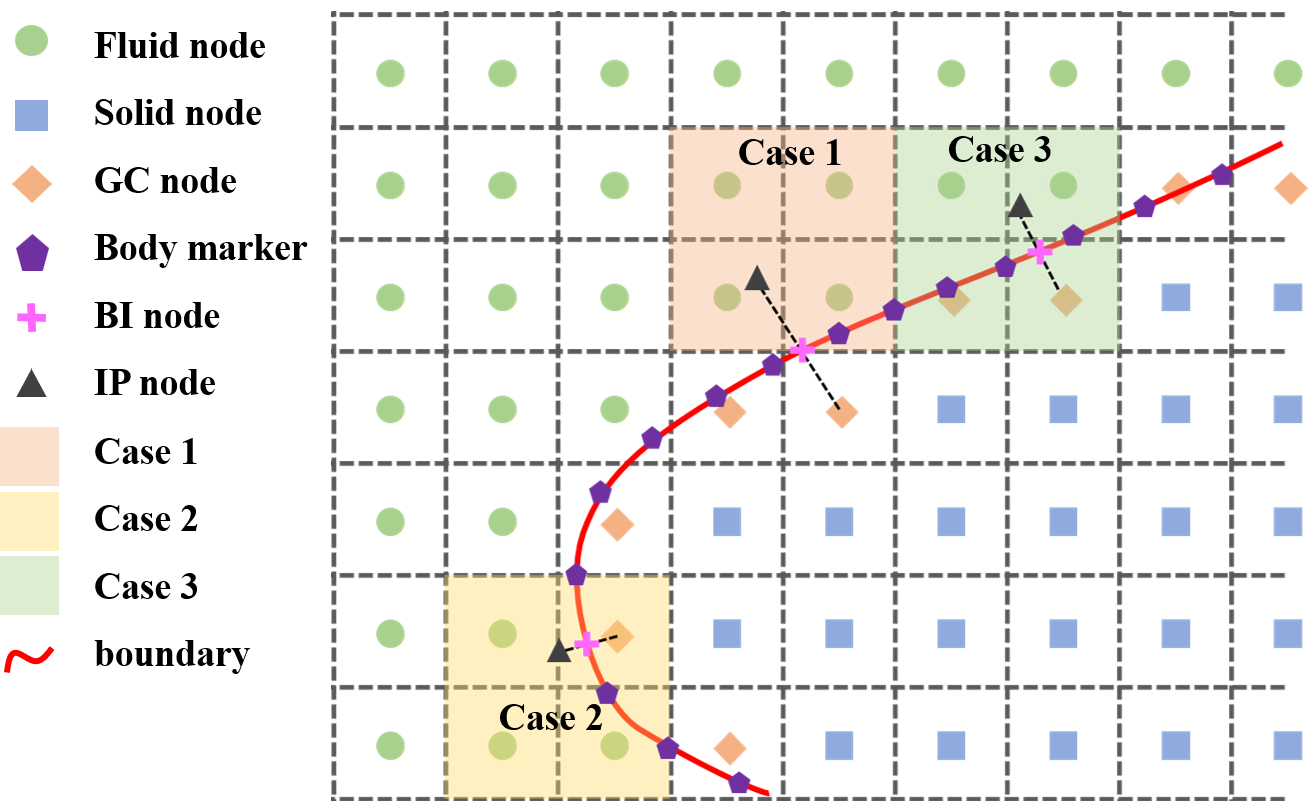}
    \caption{Schematic showing the GC-related nodes on the 2D fluid-structure boundary.}
    \label{fig2}
\end{figure}

The ghost-cell immersed boundary method (GCIBM) is employed to couple the evolution of the structure and the flow fluid.
The schematic of the GCIBM method is shown in Figure \ref{fig2}.
The boundary of the solid is represented by discrete body markers (depicted by purple pentagons) connected by linear segments,
which distinguishes the fluid cells (depicted by green circles) and the solid cells (depicted by blue squares). 
The ghost cells (depicted by orange diamonds) are virtual points within the solid domain near the fluid-structure interface, facilitating the application of boundary conditions. Imaginary point (IP) nodes (depicted by black triangles) are symmetrically located across the boundary relative to the GC nodes, while boundary intersection (BI) nodes (depicted by magenta plus signs) are defined as the midpoints between the GC and their corresponding IP nodes.

The key to GCIBM lies in calculating the pseudo flow variables in the GCs, which play a dual role in the simulation.
On one hand, they are integral to flow simulation by enforcing the no-slip and no-penetration boundary conditions. 
On the other hand, variables defined at GCs, e.g., pressure, are used to compute the forces exerted on the solid boundary, which in turn serve as boundary conditions for PD, which is crucial for capturing fluid-structure interactions.

\subsubsection{Identification of imaginary points and boundary intersections}  

After the determination of fluid and solid cells according to a given geometric shape, the location of IP and nodes can be determined via
\begin{equation}  
{\bf{x}}_{\rm{IP}} = {\bf{x}}_{\rm{GC}} + 2d {\bf{n}}_{\rm{s}}, \quad   
{\bf{x}}_{\rm{BI}} = \frac{{\bf{x}}_{\rm{IP}} + {\bf{x}}_{\rm{GC}}}{2},  
\label{eq:node_relationships}  
\end{equation} 
where \( {\bf{x}}_{\rm{GC}} \), \( {\bf{x}}_{\rm{IP}} \), and \( {\bf{x}}_{\rm{BI}} \) are the positions of the GC, IP, and BI nodes, respectively. 
Here, \( d \) is the shortest distance from the ghost cell to the boundary, and \( {\bf{n}}_{\rm{s}} \) is the outward unit normal vector at the boundary.

\subsubsection{Calculation of variables at boundary intersections}  
The BI nodes are typically positioned between boundary nodes (with possible overlap). The variables at the BI nodes can be computed using the Lagrange interpolation,
\begin{equation}  
\phi_{\mathrm{BI}} = \frac{\left( x_{\mathrm{BI}} - x_2 \right)\left( y_{\mathrm{BI}} - y_2 \right)}{\left( x_1 - x_2 \right)\left( y_1 - y_2 \right)}\phi_1   
+ \frac{\left( x_1 - x_{\mathrm{BI}} \right)\left( y_1 - y_{\mathrm{BI}} \right)}{\left( x_1 - x_2 \right)\left( y_1 - y_2 \right)}\phi_2,  
\label{eq:bi_interpolation}  
\end{equation}  
where \( (x_1, y_1) \) and \( (x_2, y_2) \) denote the two neighboring boundary nodes.

\subsubsection{Calculation of variables at imaginary points}

The bi-linear interpolation method is used to compute any flow variable \(\phi\) (e.g., pressure, velocity, and temperature) at the IP node.
Such interpolation technique utilizes the information from the nearest neighboring fluid cells around the IP. The variables \(\phi\) at the IP are expressed as
\begin{equation}
\phi = w_1 + w_2 x_{\text{IP}} + w_3 y_{\text{IP}} + w_4 x_{\text{IP}} y_{\text{IP}},
\label{eq:ip interpolation}
\end{equation}
where \(w_i \ (i = 1, 2, 3, 4)\) are the interpolation coefficients, determined by the following Eq.(\ref{eq17}) to (\ref{eq21}) based on the indices of the neighboring fluid nodes. 

If all four neighboring nodes are fluid nodes (Case 1 in Figure \ref{fig2}), the coefficients \(w_i\) are determined by:
\begin{equation}
\label{eq17}
\begin{bmatrix}
w_1 \\ w_2 \\ w_3 \\ w_4
\end{bmatrix}
=
\begin{bmatrix}
1 & x_1 & y_1 & x_1 y_1 \\
1 & x_2 & y_2 & x_2 y_2 \\
1 & x_3 & y_3 & x_3 y_3 \\
1 & x_4 & y_4 & x_4 y_4
\end{bmatrix}^{-1}
\begin{bmatrix}
\phi_1 \\ \phi_2 \\ \phi_3 \\ \phi_4
\end{bmatrix},
\end{equation}
where $\{x_{i}, y_{i}\}_{i = 1, 2, 3, 4}$ represents the locations of four neighboring fluid nodes around the IP, respectively.

If one of the four nodes belongs to the solid domain (Case 2 in Figure \ref{fig2}), the BI node replaces the solid node in the interpolation. Given the Dirichlet fluid-solid boundary condition, the coefficients \(w_i\) are calculated as
\begin{equation}
\begin{bmatrix}
w_1 \\ w_2 \\ w_3 \\ w_4
\end{bmatrix}
=
\begin{bmatrix}
1 & x_1 & y_1 & x_1 y_1 \\
1 & x_2 & y_2 & x_2 y_2 \\
1 & x_3 & y_3 & x_3 y_3 \\
1 & x_{\text{BI}} & y_{\text{BI}} & x_{\text{BI}} y_{\text{BI}}
\end{bmatrix}^{-1}
\begin{bmatrix}
\phi_1 \\ \phi_2 \\ \phi_3 \\ \phi_{\text{BI}}
\end{bmatrix},
\end{equation}
where \((x_{\text{BI}}, y_{\text{BI}})\) and \(\phi_{\text{BI}}\) represent the coordinates and interpolated values at the BI point, respectively.
Given the Neumann-type boundary condition, where the flux is specified as \( \hat{\mathbf n} \cdot \nabla \phi = h \), the bi-linear interpolation weights \(w_i\) (\(i = 1, 2, 3, 4\)) are computed by:
\begin{equation}
\begin{bmatrix}
w_1 \\ w_2 \\ w_3 \\ w_4
\end{bmatrix}
=
\begin{bmatrix}
1 & x_1 & y_1 & x_1 y_1 \\
1 & x_2 & y_2 & x_2 y_2 \\
1 & x_3 & y_3 & x_3 y_3 \\
0 & n_{x,{\rm BI}} & n_{y,{\rm BI}} & x_{\rm BI} n_{y,{\rm BI}} + y_{\rm BI} n_{x,{\rm BI}}
\end{bmatrix}^{-1}
\begin{bmatrix}
\phi_1 \\ \phi_2 \\ \phi_3 \\ h
\end{bmatrix},
\end{equation}
where \( \hat{\mathbf n} = (n_{x,{\rm BI}}, n_{y,{\rm BI}}) \) is the unit normal vector at the BI node, and \(n_{x,{\rm BI}}\) and \(n_{y,{\rm BI}}\) are its components along the \(x\) and \(y\) directions.

If two of the four nodes belong to the solid domain (Case 3 in Figure \ref{fig2}), the boundary intersection points (BI1 and BI2) replace the two solid nodes. Under the Dirichlet boundary condition, the interpolation coefficients \(w_i\) are determined by
\begin{equation}
\begin{bmatrix}
w_1 \\ w_2 \\ w_3 \\ w_4
\end{bmatrix}
=
\begin{bmatrix}
1 & x_1 & y_1 & x_1 y_1 \\
1 & x_2 & y_2 & x_2 y_2 \\
1 & x_{\text{BI1}} & y_{\text{BI1}} & x_{\text{BI1}} y_{\text{BI1}} \\
1 & x_{\text{BI2}} & y_{\text{BI2}} & x_{\text{BI2}} y_{\text{BI2}}
\end{bmatrix}^{-1}
\begin{bmatrix}
\phi_1 \\ \phi_2 \\ \phi_{\text{BI1}} \\ \phi_{\text{BI2}}
\end{bmatrix},
\end{equation}
where \((x_{\text{BI1}}, y_{\text{BI1}})\), \((x_{\text{BI2}}, y_{\text{BI2}})\), \(\phi_{\text{BI1}}\), and \(\phi_{\text{BI2}}\) represent the coordinates and interpolated variables at the two BI points.
For the Neumann-type boundary condition, where the fluxes at BI1 and BI2 are specified as \(g_1\) and \(g_2\), respectively, the interpolation coefficients are computed by
\begin{equation}
\label{eq21}
\begin{bmatrix}
w_1 \\ w_2 \\ w_3 \\ w_4
\end{bmatrix}
=
\begin{bmatrix}
1 & x_1 & y_1 & x_1 y_1 \\
1 & x_2 & y_2 & x_2 y_2 \\
0 & n_{x1,{\rm BI1}} & n_{y1,{\rm BI1}} & x_{\rm BI1} n_{y,{\rm BI1}} + y_{\rm BI1} n_{x1,{\rm BI1}} \\
0 & n_{x2,{\rm BI2}} & n_{y2,{\rm BI2}} & x_{\rm BI2} n_{y,{\rm BI2}} + y_{\rm BI2} n_{x2,{\rm BI2}}
\end{bmatrix}^{-1}
\begin{bmatrix}
\phi_1 \\ \phi_2 \\ g_1 \\ g_2
\end{bmatrix}.
\end{equation}
The obtained interpolation weights \(w_i\) are then used to compute the variables at the IP nodes using Eq.(\ref{eq:ip interpolation}).

\subsubsection{Calculation of variables in ghost cells}
\begin{figure}[htb!]
    \centering
    \includegraphics[width=0.6\textwidth]{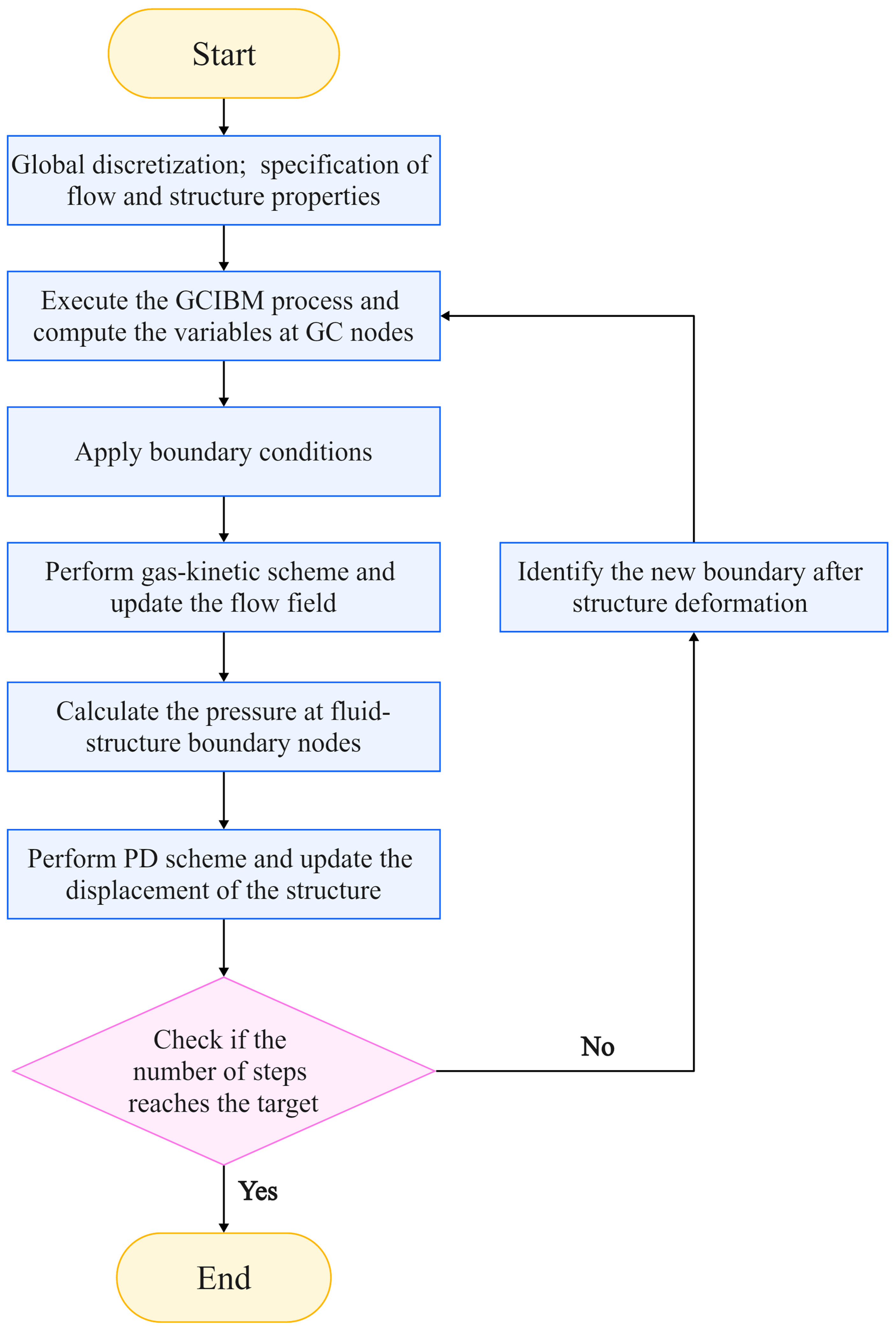} 
    \caption{The solution algorithm of fluid-structure coupling.}
    \label{fig:flowchart}
\end{figure}

Once the variables at the IP and BI nodes are determined corresponding to the given boundary conditions, the variables in the ghost cells can be calculated. 
For an arbitrary variable \(\phi\), governed by the Dirichlet boundary condition, the variables at GC nodes are calculated as 
\begin{equation}
    \phi_{\mathrm{GC}} = 2\phi_{\mathrm{BI}} - \phi_{\mathrm{IP}}.
\end{equation}
For pressure \( p \), which is governed by the Neumann boundary condition, the status in the GC is given by \(p_{\mathrm{GC}} = p_{\mathrm{BI}}\).
And thus, using the ideal gas law, the density at the GC nodes is computed as \(\rho_{\mathrm{GC}} = \frac{p_{\mathrm{GC}}}{R T_{\mathrm{GC}}}\), where \( R \) is the gas constant.
The solution algorithm of fluid-structure coupling is summarized in Figure \ref{fig:flowchart}.

\section{Numerical Experiments}\label{sec4}
\subsection{Validation of compressible flow solver} 
We first validate the gas-kinetic compressible flow solver along with the immersed boundary method.
Two benchmark cases are employed as numerical experiments, i.e., the laminar flow around a NACA0012 airfoil and the supersonic flow around a circular cylinder. 
\subsubsection{Subsonic flow around a NACA0012 airfoil}
The NACA0012 airfoil is considered at a zero angle of attack (\(\alpha = 0^\circ\)), with the freestream Mach number \(\mathrm{Ma} = 0.5\) and the Reynolds number \(\mathrm{Re} = 5000\). 
The computational domain spans \((-10D, 20D) \times (-10D, 10D)\), with the leading edge of the airfoil located at the origin. 

\begin{figure}[htb!]
    \centering
    \includegraphics[width=0.6\textwidth]{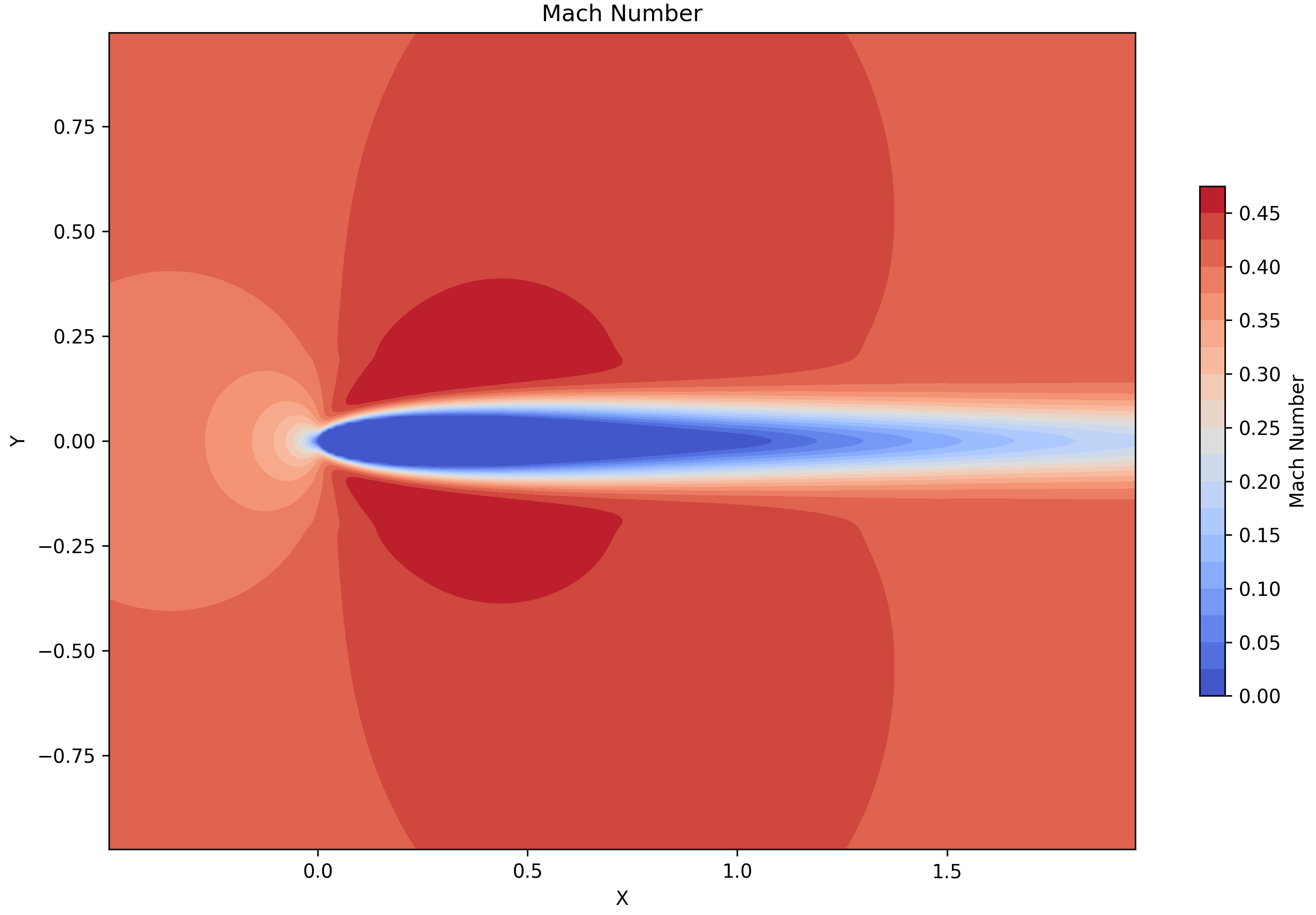}
    \caption{Mach number distribution in the flow field around the airfoil.}
    \label{fig4}
\end{figure}

\begin{figure}[htb!]
    \centering
    \includegraphics[width=0.6\textwidth]{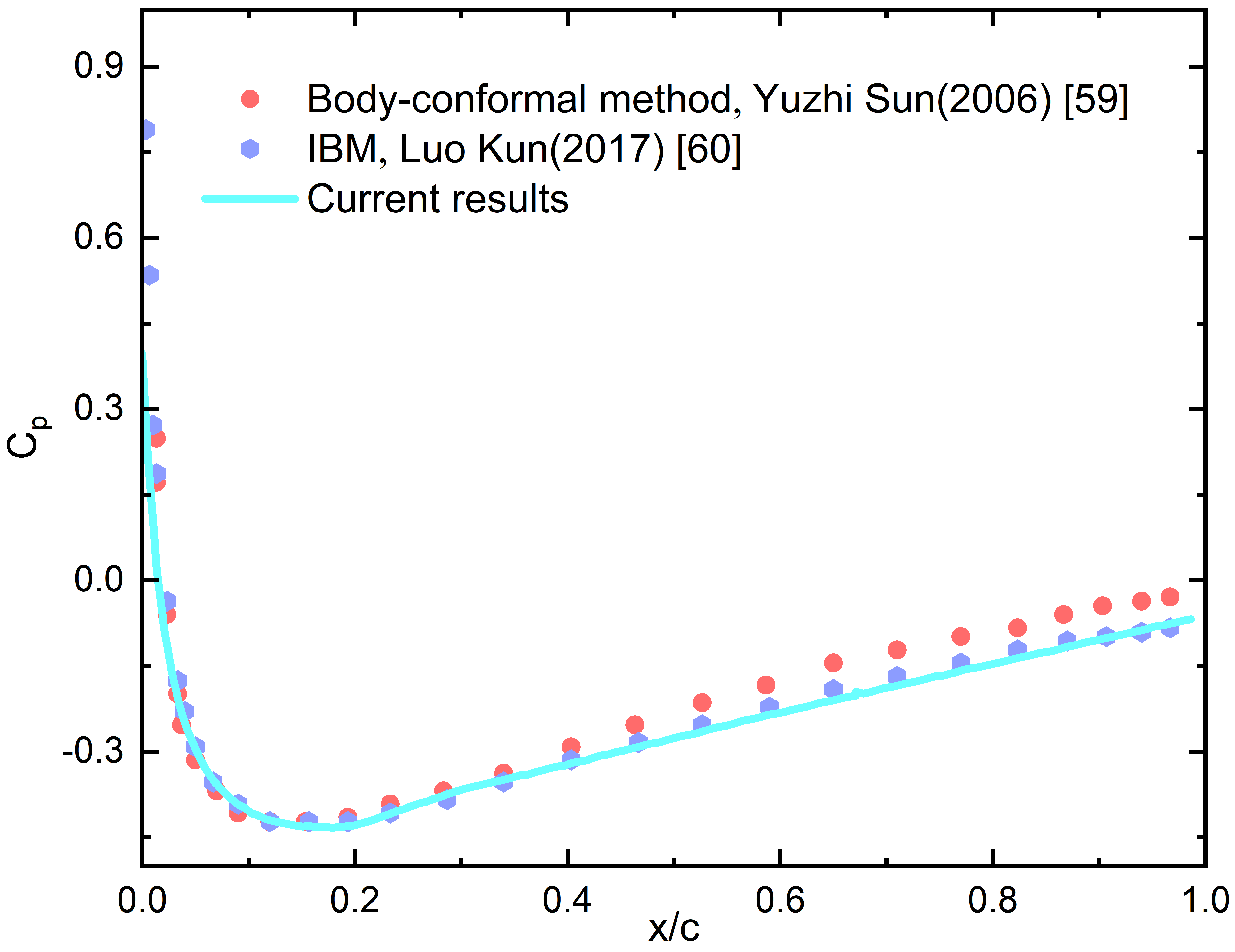}
    \caption{Comparison of the pressure coefficient distribution $C_p$ along the airfoil surface, where the horizontal axis $x/c$ represents the relative horizontal position along the airfoil, where $c$ is the airfoil chord length, and the vertical axis $C_p$ represents the pressure distribution.}
    \label{fig5}
\end{figure}

A non-uniform Cartesian grid is employed. A fine grid (\(\Delta x = \Delta y = D/512\)) is used near the airfoil, while a coarser grid (\(\Delta x = \Delta y = D/2\)) is applied in more distant areas.
Figure \ref{fig4} shows the Mach number distribution around the airfoil.
Figure \ref{fig5} presents the comparison of the pressure coefficient distribution along the airfoil surface between the current results and those from the literature \cite{sun, naca}.
The consistency between these results confirms the sufficiency of the current approach in accurately predicting viscous compressible flows with boundary effects.

\subsubsection{Supersonic flow around a circular cylinder}  

The supersonic flow around a circular cylinder is considered.
The freestream Mach number is set as \(\mathrm{Ma} = 2.0\), and the Reynolds number, \(\mathrm{Re} = 300\). The computational domain spans (0,10$D$) × (0,12$D$), with the cylinder center located at (2$D$,6$D$).

\begin{figure}[htb!]
    \centering
    \includegraphics[width=1.0\textwidth]{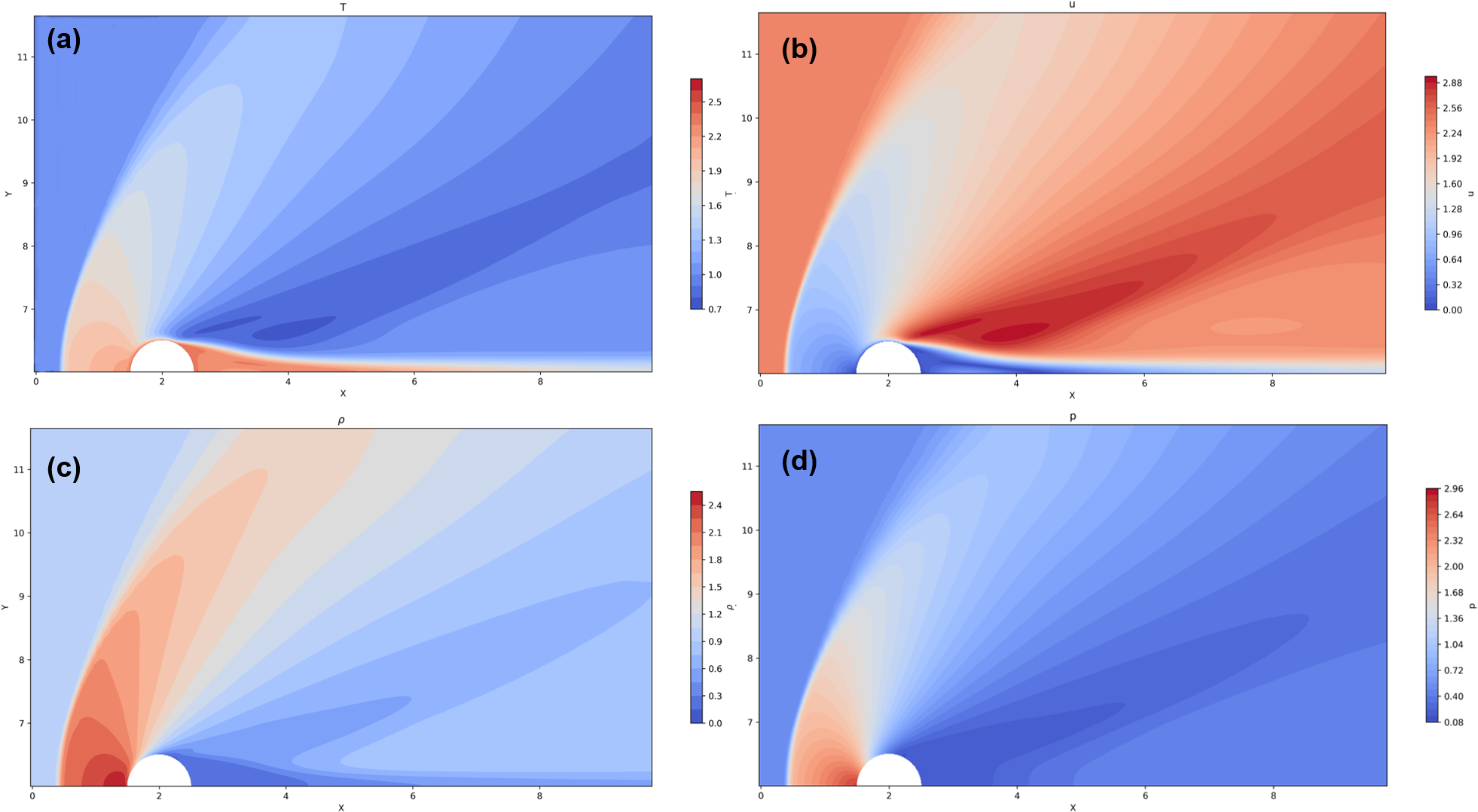}
    \caption{Distribution of solution fields in the supersonic flow around circular cylinder. (a) temperature; (b) velocity; (c) density; (d) pressure.}
    \label{fig6}
\end{figure}

\begin{figure}[htb!]
    \centering
    \includegraphics[width=0.6\textwidth]{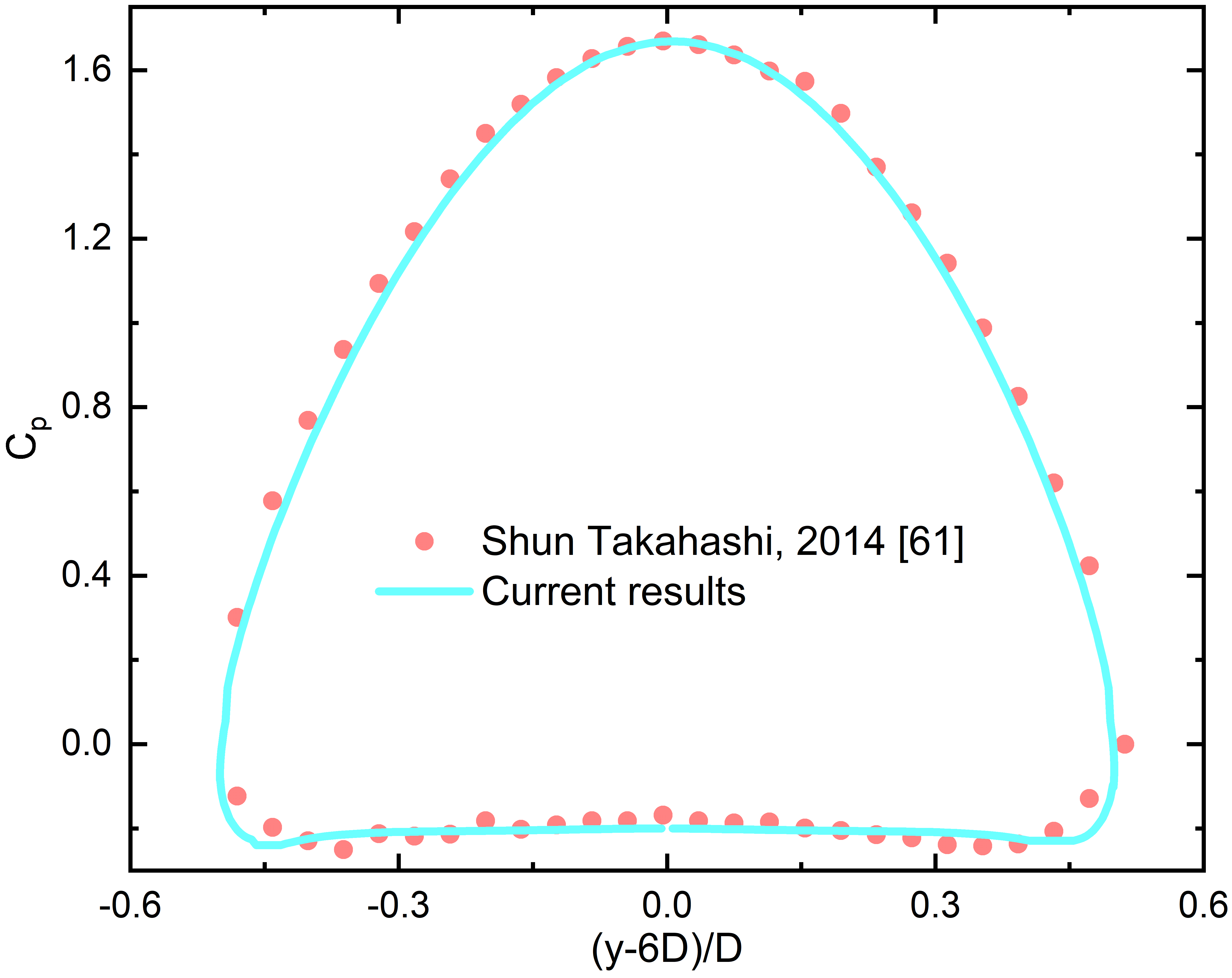}
    \caption{Comparison of the pressure coefficient distribution $C_p$ along the cylinder surface between the results from the current model and those from the literature. In this figure, the horizontal axis (y-6D)/D represents the relative vertical position along the cylinder, where D is the diameter of the cylinder, and the vertical axis $C_p$ represents the pressure distribution.}
        \label{fig7}
\end{figure}

Similar as the airfoil case, a non-uniform grid is employed. A fine grid (\(\Delta x = \Delta y = D/512\)) is used near the cylinder surface, while a coarser grid (\(\Delta x = \Delta y = D/2\)) is applied in the distant regions. 
Figure \ref{fig6} shows the Mach number distribution around the cylinder, and Figure \ref{fig7} presents a comparison of the pressure coefficient distribution along the cylinder surface between the results from the current model and those from the literature \cite{shun}.
The comparison of drag coefficients $C_D$ from different numerical methods is shown in Table \ref{tab:drag_coefficient_comparison}.
The numerical experiments demonstrate that the current approach is capable of accurate prediction of the behavior of supersonic flow as it interacts with solid boundaries.

\begin{table}[ht]
\caption{Comparison of drag coefficient predicted by  the current simulations with the reported data.}
\label{tab:drag_coefficient_comparison}
\centering
\begin{tabular}{lccc}
\toprule
\textbf{Method} & \textbf{Body-conformal\cite{sun}} & \textbf{IBM\cite{shun}} & \textbf{Current Simulation} \\
\midrule
\textbf{$C_D$} & 1.547 & 1.525 & 1.556 \\
\bottomrule
\end{tabular}
\end{table}

\subsection{Validation of peridynamic solver}

We then validate the peridynamic model for structural deformation and crack propagation.
A two-dimensional plate with an inherent crack is settled. 
The plate dimensions are \(L = 50 \, \mathrm{mm}\) (length) and \(W = 50 \, \mathrm{mm}\) (width).
A horizontal crack of length \(D = 10 \, \mathrm{mm}\) is initialized at the plate's center. 
The material is assumed to be isotropic and linear elastic with Young’s modulus $E$ of 192GPa and the density of 8000 kg/$m^3$.

\begin{figure}[htb!]
    \centering
    \includegraphics[width=0.46\textwidth]{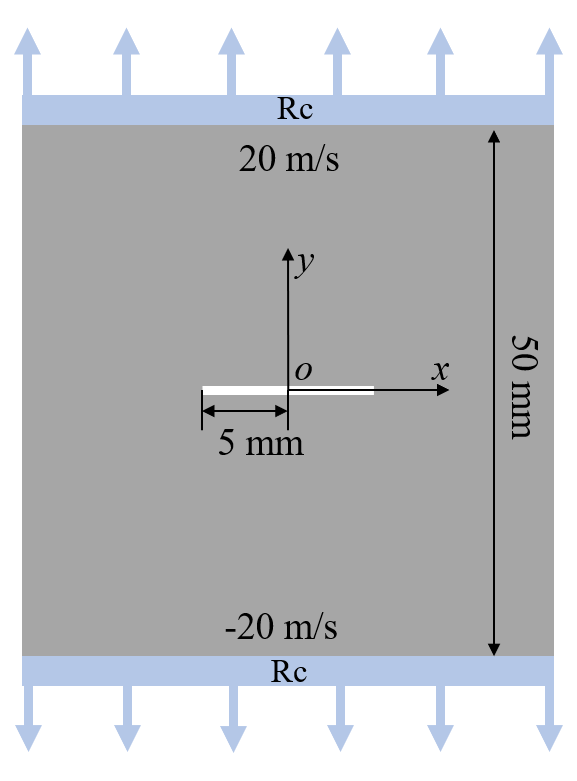}
    \caption{Schematic of the plate with a pre-existing crack.}
    \label{fig8}
\end{figure}

The plate is discretized into \(500 \times 500\) material points.
The velocity boundary conditions of $u_y = \pm 20 \, \mathrm{m/s}$ are applied along its upper and lower edges. 
Two fictitious boundary layers, \(R_c\), each containing \(500 \times 3\) material points, are added at the top and bottom to enforce the boundary condition. 
The simulation employs a time step of \(\Delta t = 1.3367 \times 10^{-8} \, \mathrm{s}\).
The schematic of the numerical experiment is shown in Figure \ref{fig8}.

\begin{figure}[htb!]
    \centering
    \includegraphics[width=0.6\textwidth]{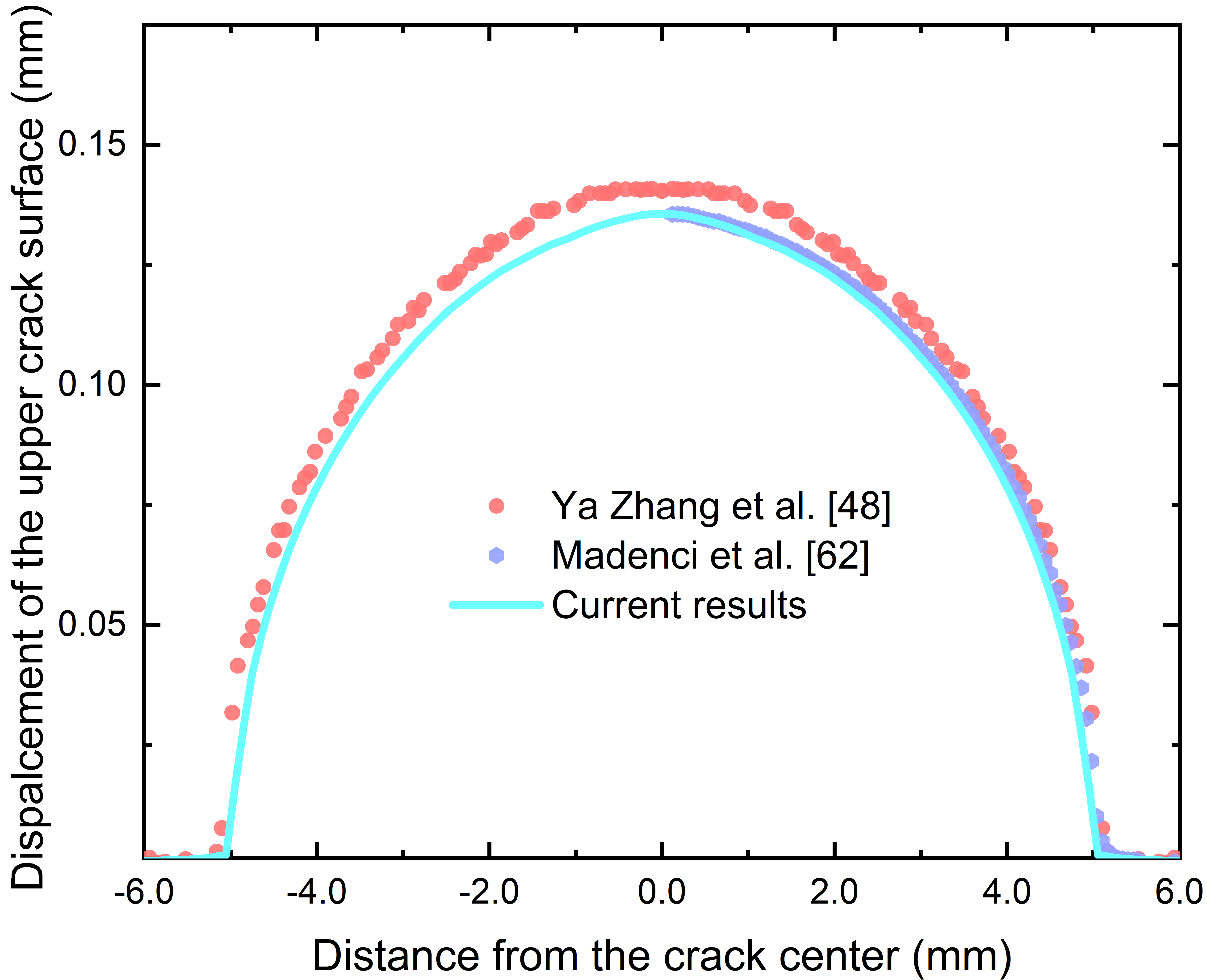}
    \caption{Vertical displacement of the material points near the upper surface of initial crack at the time t = 16.7 $\mu s$ when failure is not allowed.}
    \label{fig9}
\end{figure}

\begin{figure}[htb!]
    \centering
    \includegraphics[width=0.6\textwidth]{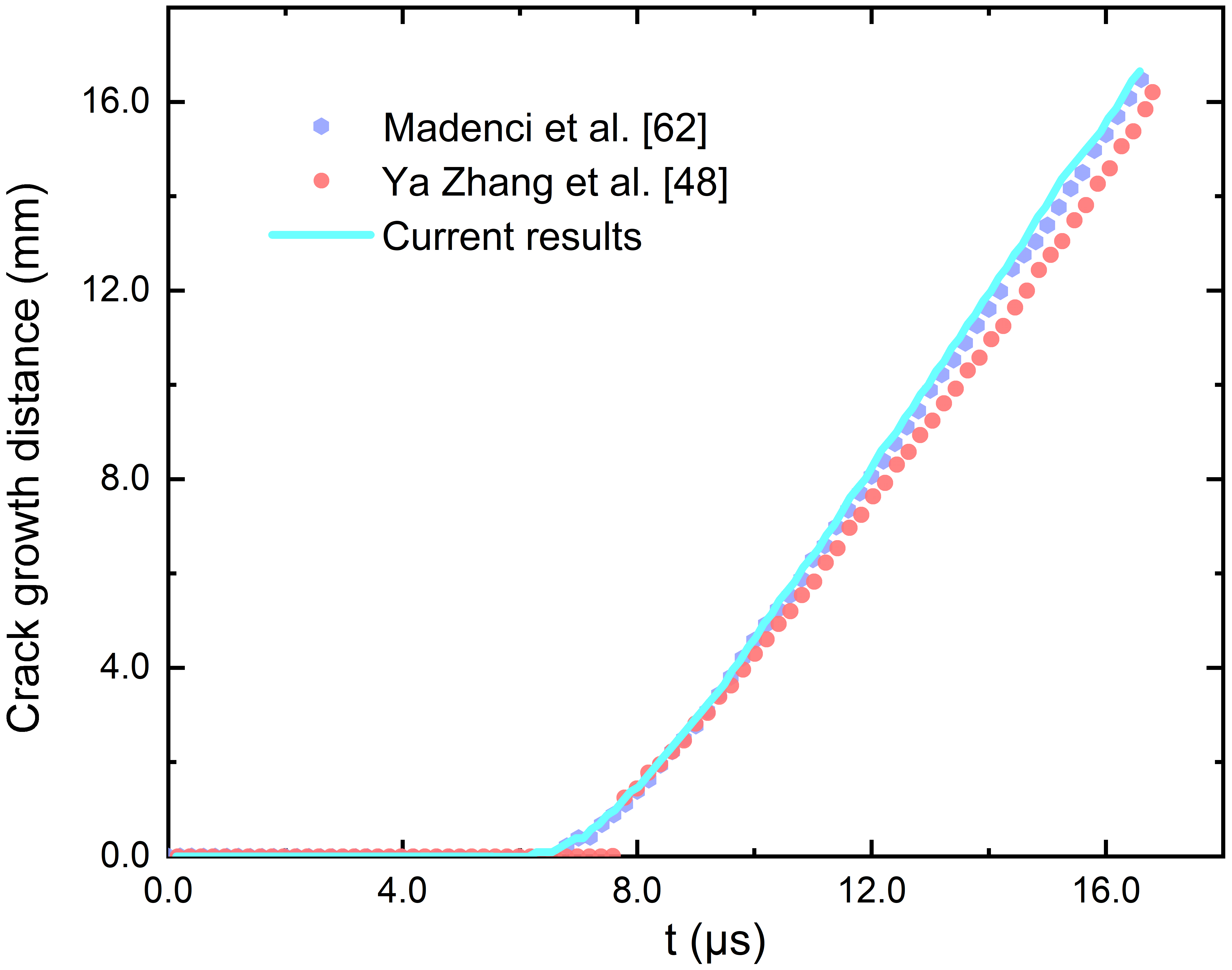}
    \caption{Variation in the propagation distance of the right crack tip during growth, when failure occurs.}
    \label{fig10}
\end{figure}

The fracture behavior of the plate is governed by the critical stretch \( s_c \).
Figure \ref{fig9} presents the comparison of the shape of cracks predicted by the current method and literature results at \( s_c = 1.0 \).
Figure \ref{fig10} provides the predicted fracture propagation velocity compared with the reference solution \cite{RN36, Madenci2014}.
The overall good agreement suggests that the current bond-based peridynamic solver can describe material deformation and damage.

\subsection{Validation of the coupling scheme}
\begin{figure}[htb!]
    \centering
    \includegraphics[width=0.75\textwidth]{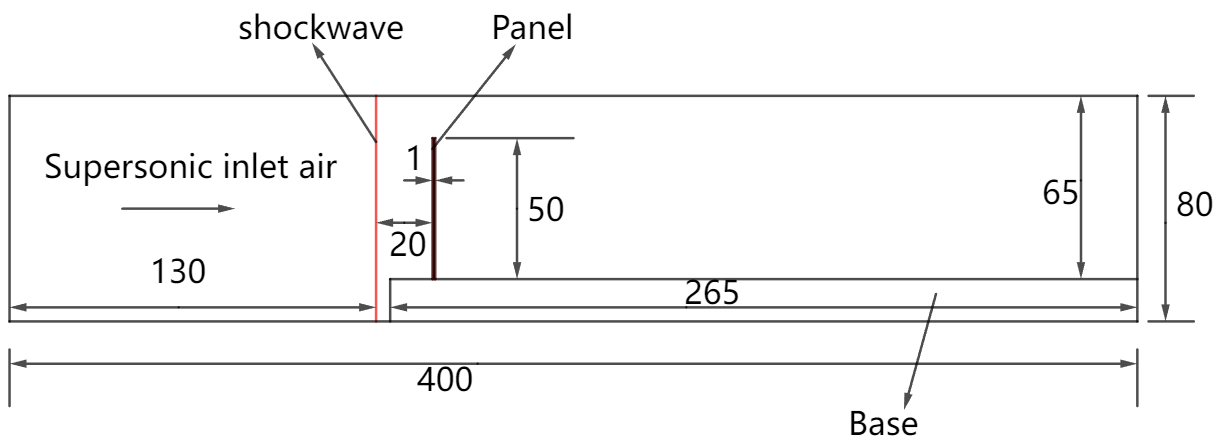}
    \caption{Schematic of the shock tube experiment.}
    \label{fig11}
\end{figure}

Here, the fluid-structure interaction solver is to be validated using the experimental setup by Giordano \cite{wave}. The Giordano model features an elastic panel fixed at its lower edge to a rigid base. This setup is placed in a shock tube consisting of a movable high-pressure chamber (0.75 m long), a fixed low-pressure chamber (2.02 m long), and a movable experimental chamber (0.98 m long), as summarized in Figure \ref{fig11}. 

\begin{figure}[htb!]
    \centering
    \includegraphics[width=0.8\textwidth]{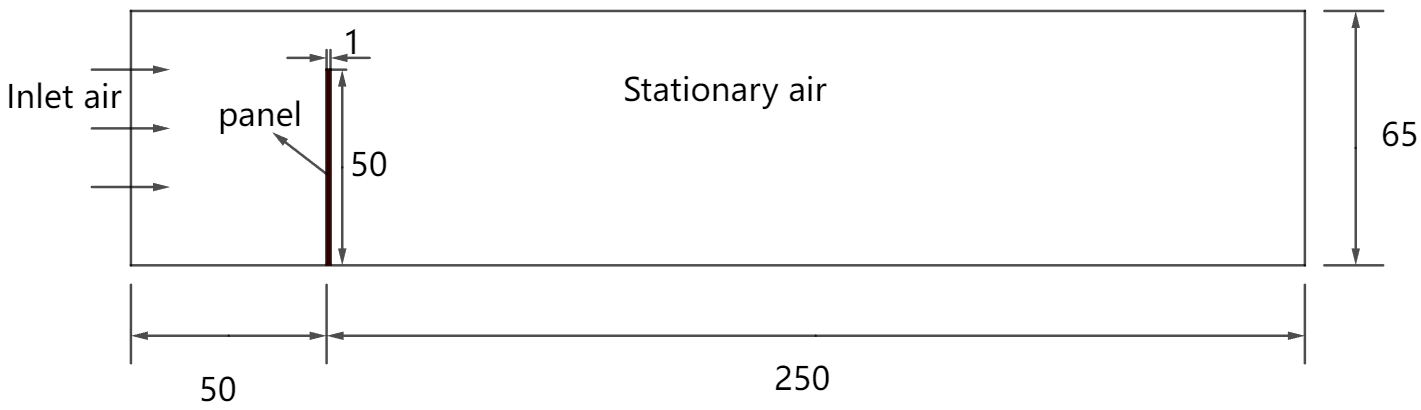}
    \caption{Simplified geometric configuration for numerical simulation.}
    \label{fig12}
\end{figure}

Ning et al.\cite{com2} simplified the model by assuming a tube length of 300 mm with a positive shock imposed at the left boundary. The base was removed under the assumption that its impact on the results is negligible. In the simplified model, the panel has a length of 50 mm and a thickness of 1 mm, and its material behavior is linear elastic. The simplified experimental setup is illustrated in Figure \ref{fig12}. The initial flow field and material parameters are summarized in Table \ref{tab1}. The material of panel is assumed to be isotropic and linear elastic with Young’s modulus $E$ of 220 GPa and the density of 7850 kg/$m^3$. 

\begin{table}[ht]
\centering
\caption{Fluid properties and flow conditions for the simulation.}
\label{tab1}
\begin{tabular}{cp{2cm}p{2cm}p{2.4cm}p{2cm}} 
\toprule
\textbf{Material} & \textbf{Density (kg/m\(^3\))} & \textbf{Pressure (Pa)} & \textbf{Specific Heat Ratio} & \textbf{Velocity (m/s)} \\ 
\midrule
Stationary air & 1.20 & 101.0 & 1.4 & 0 \\ 
Inflow air & 1.63 & 155.7 & 1.4 & 109.6 \\ 
\bottomrule
\end{tabular}
\end{table}

The computational domain spans $(0, 6L) \times (0, 1.3L)$. To reduce computational cost, a non-uniform grid is used in the $x$-direction, with finer grid spacing ($\Delta x = 0.002L$) near the plane and coarser grid spacing ($\Delta x = 0.02L$) farther away. Since the plane length is comparable to the domain size in the $y$-direction, a uniform grid is adopted in the $y$-direction, with a grid spacing of $\Delta y = 0.02L$.

\begin{figure}[htb!]
    \centering
    \includegraphics[width=0.8\textwidth]{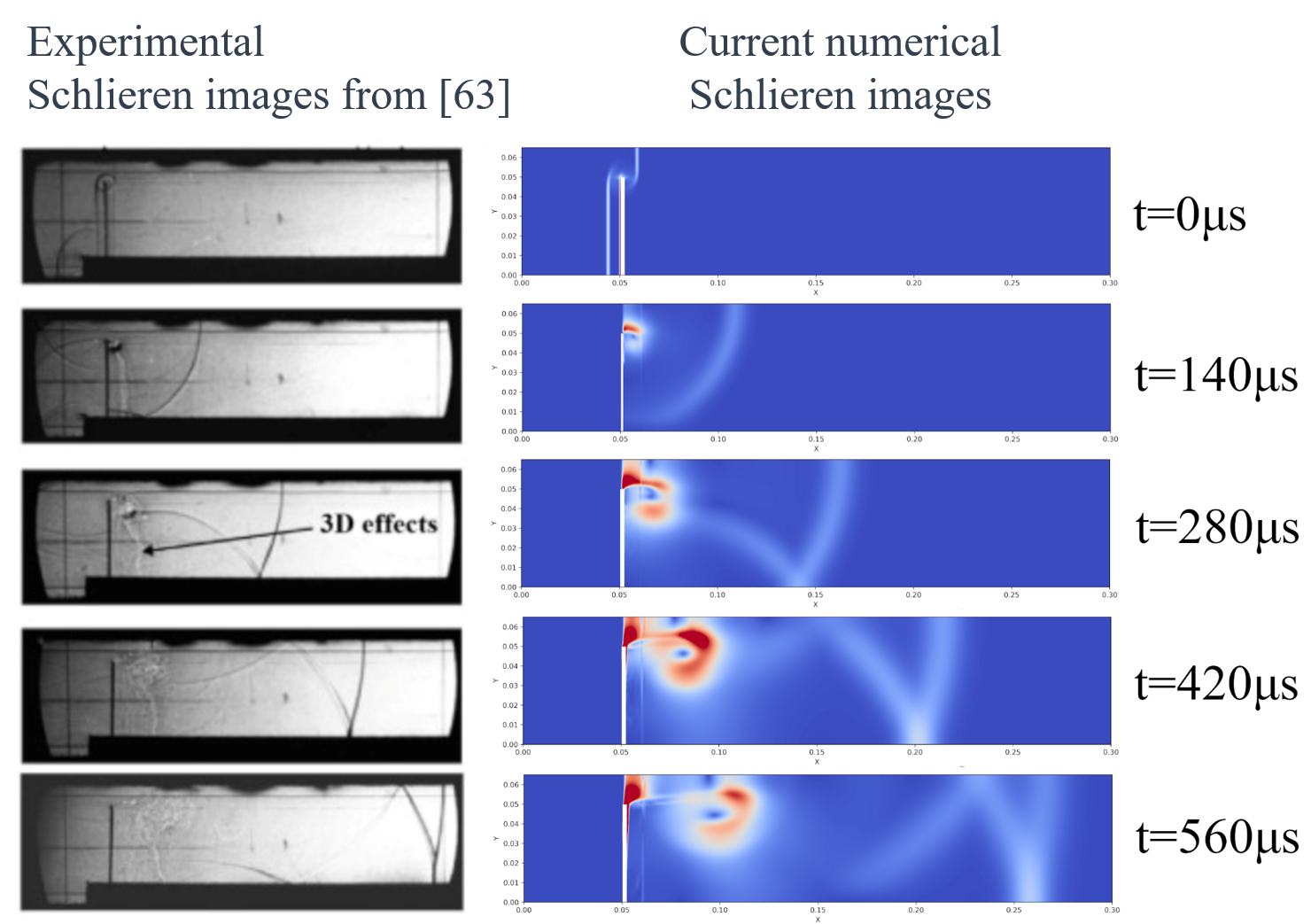}
    \caption{Comparison between the time evolution of (a) the experimental Schlieren photos from \cite{wave} and (b) our numerical results for the panel with 50 mm length.}
        \label{fig13}
\end{figure}

\begin{figure}[htb!]
    \centering
    \includegraphics[width=0.7\textwidth]{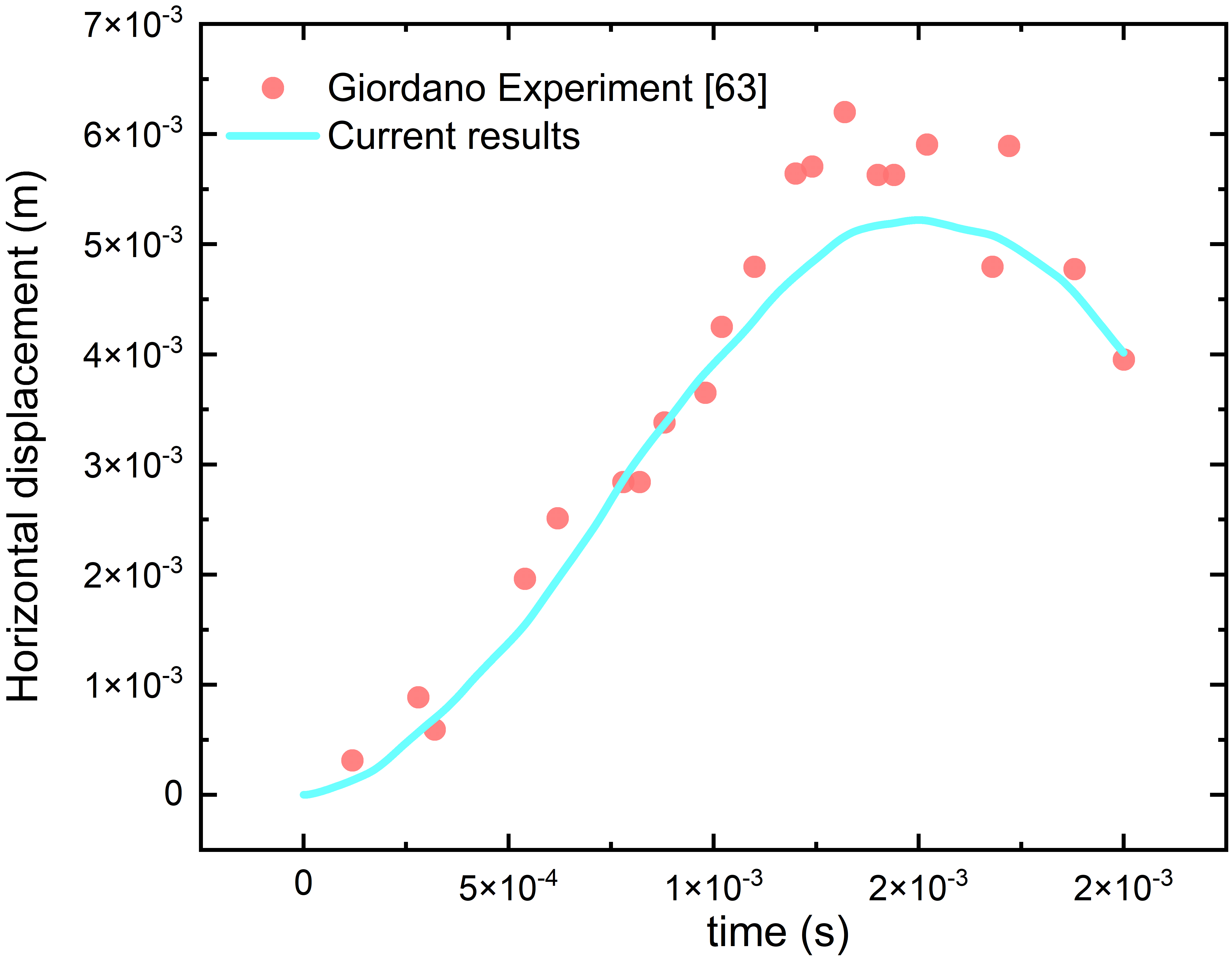}
    \caption{Variation in the horizontal displacement of the panel end, where the experiment data refer to Giordano \cite{wave}.}
        \label{fig14}
\end{figure}

Figure \ref{fig13} shows the comparison of the Schlieren images between \cite{wave} and our simulation results. Figure \ref{fig14} shows the displacement of the panel's end during the loading process over 2 ms, based on both the experimental and current simulation data. The numerical results show good agreement with the experimental data \cite{wave}. In the time range from 0 to about 1.5 ms, the displacement increase is consistent with the experimental data. After 1.5 ms, the decrease in displacement also agrees well with the experimental results.

\subsection{Fracture inside a cylinder induced by supersonic flow}
Finally, we consider a typical supersonic fluid-solid interaction scenario.
In this case, 
the free stream moving of Mach number 1.7 impacts a cylinder that contains a pre-existing internal crack.
Upon impact, the crack propagates, showcasing the distinct advantages of the current PD-based FSI model.

\begin{figure}[htb!]
    \centering
    \includegraphics[width=0.35\textwidth]{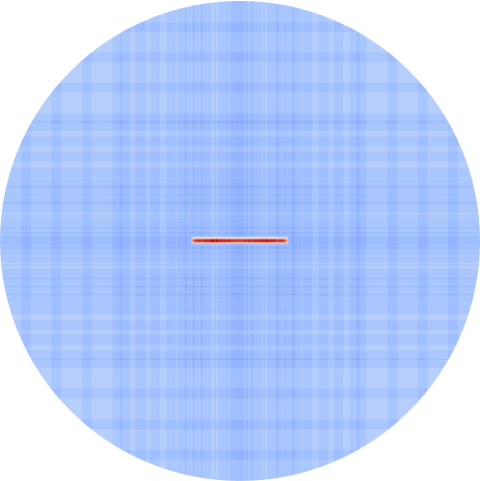}
    \caption{Pre-existing crack inside the cylinder.}
    \label{fig15}
\end{figure}

The supersonic flow with \(\mathrm{Ma}=1.7\) and \(\mathrm{Re}=200,000\) is considered in the simulation.
The computational domain spans (0,10$D$) × (0,12$D$), with the cylinder center located at (2$D$,6$D$). 
A non-uniform grid is used with a fine grid (\(\Delta x = \Delta y = D/128\)) near the cylinder and a coarser grid (\(\Delta x = \Delta y = D/2\)) in the distant regions.
The material is assumed to be isotropic and linear elastic with the density of 100 kg/$m^3$. The distance between material points is \(\Delta x = 0.005\), and
the horizon size is set as \(\delta = 4\Delta x\). An initial crack is pre-defined inside the cylinder, as shown in Figure \ref{fig15}.

\begin{figure}[htb!]
    \centering
    \includegraphics[width=0.5\textwidth]{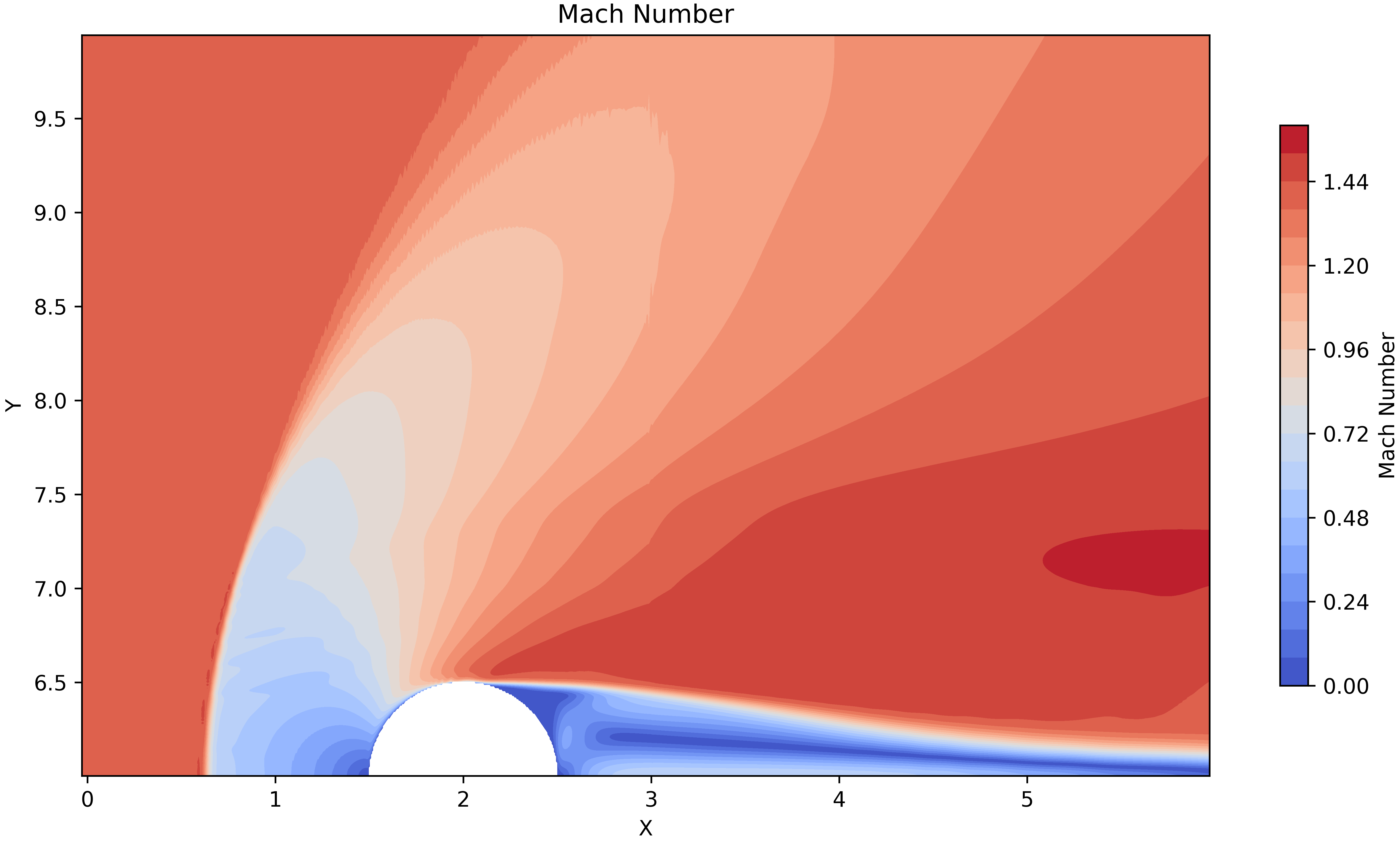}
    \caption{Mach number distribution in the flow field around the cylinder.}
    \label{fig16}
\end{figure}

\begin{figure}[htb!]
    \centering
    \includegraphics[width=0.7\textwidth]{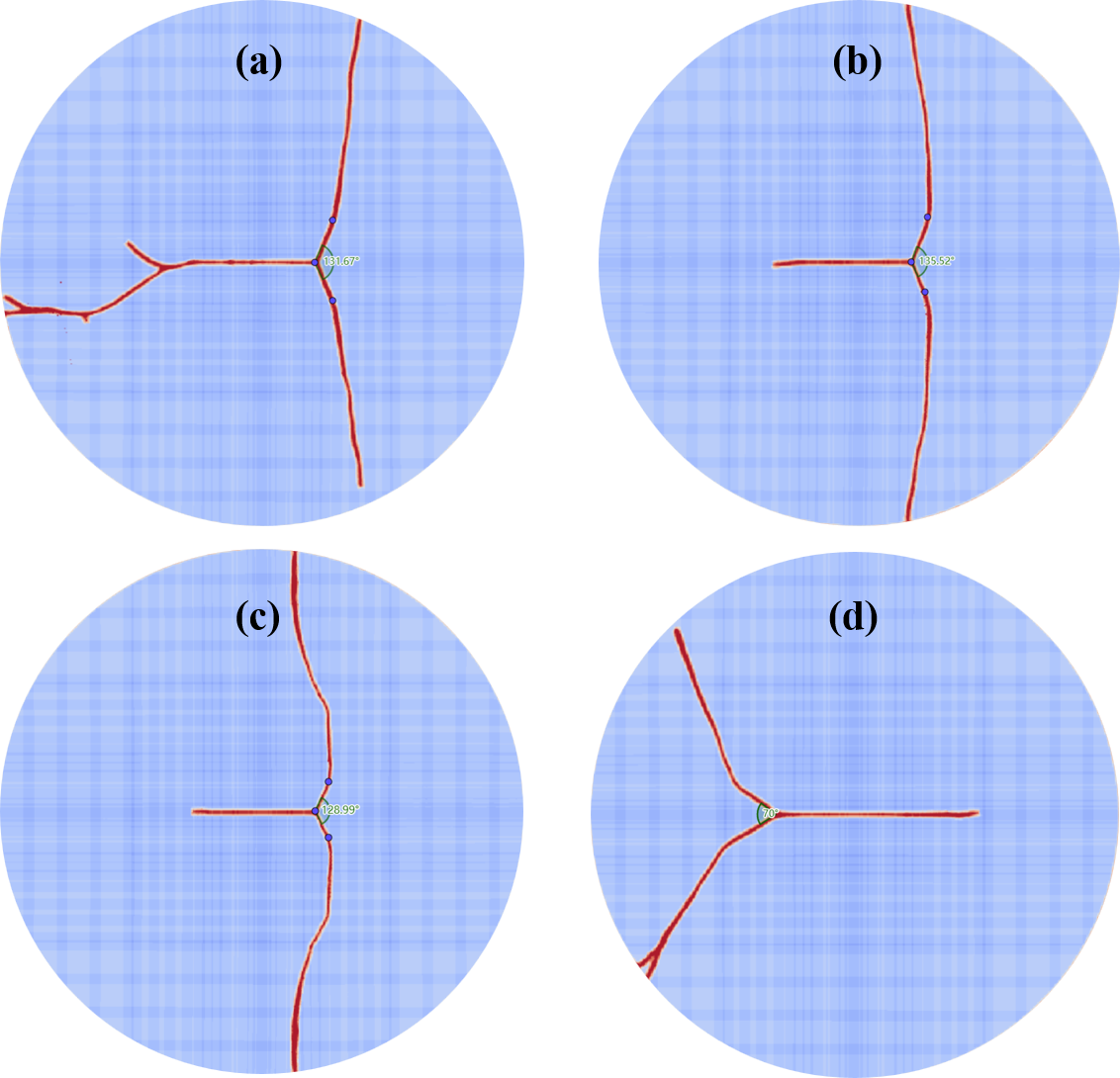}
    \caption{Fracture in the cylinder induced by a supersonic flow. The cylinders in (a), (b), (c), and (d) have the Young's moduli of 30 MPa, 40 MPa, 50 MPa, and 60 MPa, respectively.}
    \label{fig17}
\end{figure}

Figure \ref{fig16} shows the Mach number distribution around the upper half of the cylinder, and Figure \ref{fig17} illustrates crack propagation inside the cylinder under flow impact for different values of Young's modulus. For Young's moduli of 30 MPa, 40 MPa, and 50 MPa, the crack bifurcates as it propagates to the right, with a bifurcation angle of approximately 130°. The curvature of the bifurcated crack increases with the Young's modulus. However, when the Young's modulus reaches 60 MPa, the behavior reverses, and the crack bifurcates to the left, with a bifurcation angle of around 70°.
The results of numerical experiments show that different flow properties and physical parameters are decisive for the mode and rate of crack extension.
This fully justifies the need to develop coupled flow-solid solvers that can be applied to fracture mechanics.

\section{Conclusion}

In this work, we present a novel hybrid algorithm that seamlessly couples compressible fluid mechanics with deformable material behavior within a unified mesoscopic framework. 
The gas-kinetic scheme and bond-based peridynamics are utilized to characterize the dynamics of fluids and solids, respectively.
The integration of these two methods via the ghost-cell immersed boundary method facilitates a synchronized evolution of both fluid and solid fields.

The extensive numerical experiments, including subsonic and supersonic flows around canonical geometries, crack propagation, and shock wave impacts on elastic panels, demonstrate the algorithm's efficacy and its ability to predict the deformation, damage, and fracture of materials caused by high-speed flows.
These results underscore the method’s potential to address challenging fluid-structure interaction problems that are difficult to solve with the  traditional approaches.

Looking ahead, further investigations will focus on extending this framework to three-dimensional thermal interactions between fluids and structures, refining computational efficiency through adaptive mesh and time-stepping strategies, and incorporating additional physical effects to broaden the model’s applicability. 
Overall, the proposed method represents a significant step forward in the simulation of coupled fluid and solid dynamics, offering a powerful tool for both fundamental research and engineering applications.

\section*{Acknowledgment}
The current research is funded by the National Science Foundation of China (No. 12302381) and the Chinese Academy of Sciences Project for Young Scientists in Basic Research (YSBR-107).
The computing resources provided by Hefei Advanced Computing Center and ORISE Supercomputer are acknowledged.

\bibliographystyle{elsarticle-num}
\bibliography{fsi} 
\end{document}